\documentclass[11pt]{article}
\usepackage{a4}
\usepackage{times}
\usepackage{graphicx}
\usepackage{color}
\usepackage{subfig}

\begin{document}
\title{Understanding better (some) astronomical data using Bayesian methods
%
}
\author{
S. Andreon,$^1$\thanks{stefano.andreon@brera.inaf.it}
\\
$^1$INAF--Osservatorio Astronomico di Brera, Milano, Italy\\
}

\maketitle

\begin{abstract}
\noindent
Current analysis of astronomical data are confronted with the daunting task
of modeling the awkward features of astronomical 
data, among which heteroscedastic (point-dependent) 
errors, intrinsic scatter, non-ignorable data collection   
(selection effects), data structure, non-uniform populations (often
called Malmquist bias), non-Gaussian data, and upper/lower limits.
This chapter shows, by examples, how modeling all these features 
using Bayesian methods. In
short, one just need to formalize, using maths, the logical link between
the involved quantities, how the data arize and what we already known on
the quantities we want to study. The posterior probability distribution 
summarizes what we known on the studied quantities after the data, and we should
not be afraid about their
actual numerical computation, because it is left to (special) Monte Carlo 
programs such as JAGS. As examples, we show
how to predict the mass of a new object disposing of a calibrating
sample, how to constraint cosmological parameters from supernovae
data and how to check if the fitted data are in tension with the adopted
fitting model. Examples are given with their coding. These examples can be easily used
as template for completely different analysis, on totally unrelated
astronomical objects, requiring to model the same awkward data features.
\end{abstract}

\section{Introduction}

Astronomical data present a number of quite common awkward features (see
\cite{AandH11review} for a review):

\begin{itemize}

\item  
{\bf heteroscedastic errors}: error sizes vary from point to point.

\item  
{\bf non-Gaussian data}: the likelihood is asymmetric and
thus errors are, e.g $3.4^{+2.5}_{-1.2}$. 
Upper/lower limits, as 2.3 at 90 \% probability, are (perhaps extreme) 
examples of asymmetric likelihood. 

\item  
{\bf non uniform populations or data structure}:
the number of objects per unit parameter(s) is non-uniform. 
This is the source of the Malmquist- or Eddington- like bias that affect
most astronomical quantities, as parallaxes, star and galaxy counts, mass
and luminosity, galaxy cluster velocity dispersions, supernovae luminosity
corrections, etc. 

\item  
{\bf intrinsic scatter}: data often scatter more than allowed by the errors. The
extra-scatter can be due to unidentified sources of errors, often called
systematic errors, or indicates an intrinsic spread of the population
under study, i.e. the fact that astronomical objects are not identically equal.

\item  
{\bf noisy estimates of the errors}: 
as every measurement, errors are known with a finite degree of precision.
This is even more true when one is measuring complex, and somewhat model
dependent, quantities like mass.

\item  
{\bf non-random sampling}: in simple terms, the objects
in the sample are not a random sampling of those in the Universe. 
In some rare occasions in astronomy, 
sampling is planned to be non-random on purpose, but 
most of the times non-random sampling is due
to selection effects: harder-to-observe objects are very
often missed in samples. 

\item  
{\bf mixtures}: very often,
large samples include the population under interest, but also
contaminating objects. Mixtures also arise when one measure the flux
of a source in presence of a background (i.e. always).
 
\item  
{\bf prior}: we often known from past data or from
theory that some values of the parameters are more likely than other.
In order terms, we have prior knowledge about the parameter
being investigated. If we known anything, not even the order of magnitude,
about a parameter, it is  difficult even to choose which instrument,
or sample, should be used to measure the parameter.

\item  
{\bf non-linear}: laws of Nature can be more complicated than $y=ax+b$.

\end{itemize}

Bayesian methods allow to deal with these features (and also other ones), even 
all at the same time, as we illustrate in Sec 3 and 4 with two research examples, it is just
matter of stating in mathematical terms our wordy statements about the nature
of the measurement and of the objects being measured. The probabilistic (Bayesian)
approach returns the whole (posterior) probability distribution of the 
parameters, very often in form of a Monte Carlo sampling of it. 

In this paper we make an attempt to be clear at the cost of being
non-rigorous. We defer the reader looking for rigour to general textboox, 
as \cite{gelman2004bayesian}, and, to \cite{AandH10} for our first research example. 

\section{Parameter estimation in Bayesian Inference}

Before adressing a research example, let's consider an idealized applied problem
to explain the basics of the Bayesian approach.

Suppose one is interested in estimating the (log) mass of a galaxy cluster,
$lgM$.
In advance of collecting any data, we may have certain 
beliefs and expectations about the values of $lgM$.
In fact, these thoughts are often used in deciding which 
instrument will be used to gather data and how this
instrument may be configured. For example, if we plan to
measure the mass of a poor cluster via the virial theorem, 
we will select a spectroscopic
set up with adequate resolution, in order to avoid that velocity
errors are comparable to, or larger than, the likely low
velocity dispersion of poor clusters.
Crystalising these thoughts in the form of a probability distribution for
$lgM$ provides the prior $p(lgM)$, on which relies
the feasibility section of the telescope time proposal, where 
instrument, configuration and exposure time are set.

For example one may believe (e.g. from the cluster being somewhat poor)
that the log of the cluster mass is probably not far from $13$, plus or minus 1;
this might be modeled by saying that the (prior) probability distribution of the
log mass, here denoted $lgM$ is a Gaussian centred on $13$ and with $\sigma$, the 
standard deviation, equal to $0.5$, i.e. $lgM \sim \mathcal{N} (13,0.5^2)$.

Once the appropriate instrument and its set up have been selected, 
data can be collected.
In our example, this means we record a measurement of log mass, say
$obslgM200$, via, for example, a virial theorem analysis, i.e. measuring
distances and velocities. 

The likelihood describes how the noisy observation 
$obslgM200$ arises given a value of $lgM$.
For example, we may find that the measurement technique allows us to measure
masses in an unbiased way but with a standard error of 0.1 and
that the error structure is Gaussian, ie.
$obslgM200 \sim \mathcal{N} (lgM,0.1^2)$, where the tilde symbol reads ``is drawn from"
or ``is distributed as".
If we observe $obslgM200=13.3$ we usually summarise the above by writing
$lgM=13.3\pm 0.1$.

How do we update our beliefs about the unobserved log mass $lgM$ in light
of the observed measurement, $obslgM200$? 
Expressing this probabilistically, what is the posterior distribution
of $lgM$ given $obslgM200$, i.e. $p(lgM \ | \ obslgM200)$?
Bayes Theorem (\cite{Bayes}, \cite{Laplace}) tells us that
\begin{eqnarray}
p(lgM \ | \ obslgM200) \propto  p(obslgM200 \ |\ lgM) p(lgM)
\nonumber
\end{eqnarray}

i.e. the posterior (the left hand side) is equal to the product of 
likelihood and prior (the right hand side) times a proportionality 
constant of no importance in parameter estimation.

Simple algebra shows that in our example the posterior distribution of
$lgM \ | \ obslgM200$ is Gaussian, with mean
$\mu=\frac{13.0/0.5^2+13.3/0.1^2}{1/0.5^2+1/0.1^2}=13.29$
and $\sigma^2=\frac{1}{1/0.5^2+1/0.1^2}=0.0096$.
$\mu$ is just the usual weighted average of
two ``input" values, the prior and the observation, with weights
given by prior and observation $\sigma$'s. 


From a computational point of view, only simple examples
such as the one described above can generally be tackled analytically,
realistic analysis should be
instead tacked numerically by special (Markov Chain) 
Monte Carlo methods. These are included in 
BUGS-like programs (\cite{lunn2009bugs}) such as
JAGS (\cite{JAGS}), allowing scientists to focus on formalizing in mathematical terms our wordy
statements about the quantities under investigation
without worrying about the numerical implementation.
In the idealized example, we just need to write in an ascii file the symbolic
expression of the prior, $lgM \sim \mathcal{N} (13,0.5^2)$, and of
likelihood, $obslgM200 \sim \mathcal{N} (lgM,0.1^2)$
to get the posterior in form of samplings. From the
Monte Carlo sampling one may directly derive mean values, standard deviations,
and confidence regions. For example, for a 90 \% interval it is sufficient to
peak up the interval that contain 90 \% of the samplings.

\section{First example: predicting mass from a mass proxy}

Mass estimates are one of the holy grails of astronomy. Since these are
observationally expensive to measure, or 
even unmeasurable with existing facilities, 
astronomers use mass proxies, far less expensive to acquire:
from a(n usually small)
sample of objects, the researcher measures masses, $y$ and  
the mass proxy, $x$. Then, he regress $x$ vs $y$ and infer $y$ for
those objects having only $x$.
This is the way most of the times galaxy cluster masses are estimated,
for example using
the X-ray luminosity, X-ray temperature, $Y_X$, $Y_{SZ}$, the
cluster richness or the total optical luminosity. 
Here we use the cluster richness, i.e. the number of member galaxies, but
with minor changes this example can be adapted for other cases.

\cite{AandH11review} shows that predicted $y$ using the Bayesian approach 
illustrated here are more
precise than any other method and that the Bayesian approach does not show the large
systematics of other approaches. This means, in the case of masses,
that more precise masses can be derived for the same input data, i.e. at
the same telescope time cost.

\subsection{Step 1: put in formulae what you know}

\subsubsection{Heteroscedasticity}
Clusters have widely different richnesses, and thus widely different errors.
Some clusters have better determined masses than other.
Heteroscedasticity means that errors have an index
$i$, because they differ from point to point.

\subsubsection{Contamination (mixtures), non-Gaussian data and upper limits}
Galaxies in the cluster direction are both cluster members and galaxies
on the line of sight (fore/background). The contamination may be estimated by
observing a reference line of sight (fore/background), perhaps with a $C_i$ times larger
solid angle (to improve the quality of the determination).

The mathematical translation of our words, when counts are modeled
as Poisson, is:

\begin{eqnarray}
obsbkg_i \sim& \mathcal{P}(nbkg_i) &\mbox{\# Poisson with intensity
} nbkg_i \\
obstot_i \sim& \mathcal{P}(nbkg_i/C_i+n200_i) &\mbox{\# Poisson with
intensity } (nbkg_i/C_i+n200_i) 
\end{eqnarray}

The variables $n200_i$ and $nbkg_i$ represent the true richness 
and the true background galaxy counts in the studied solid angles, whereas
we add a prefix ``obs" to indicate the 
observed values. 

Upper limits are automatically accounted for. Suppose, for exposing simplicity,
that we observed 5 galaxies, $obstot_i=5$, in the cluster
direction and that in the control field direction (with $C_i=1$ for exposing
simplicity) we observe
four background galaxies, $obsbkg=4$. With one net galaxy and 
Poisson fluctuations of a few, $n200_i$ is poorly determined at best, and the
conscientious researcher would probably reports 
an upper limits of a few. To
use the information contained in the upper limit in our regression analysis we
only need to list in the data file the raw measurements ($C_i=1$, $obstot_i=5$,
$obsbkg_i=4$), as for the other
clusters. These data will be treated independently on whether an astronomer 
decides to report a measurement or an upper limit, because Nature doesn't care
about astronomer decisions.

\subsubsection{Non-linearity and extra-scatter}

The relation between mass, $M200$, and proxy, $n200$, (richness) is usually parametrized as
a power-law:

\begin{eqnarray}
M200_i &\propto& n200_i^\beta \nonumber
\end{eqnarray}

Allowing for a Gaussian intrinsic scatter $\sigma_{scat}$ 
(clusters of galaxies of a given richness may not all have the very same
mass) and taking the log, previous equation becomes:

\begin{eqnarray}
lgM200_i \sim & \mathcal{N}(\alpha+\beta \log(n200_i),
\sigma_{scat}^2) &\mbox{\# Gaussian scatter around}
(M200_i \propto n200_i^\beta) 
\label{eqn:eqn13}
\end{eqnarray}

where the intercept is $\alpha$ and the slope is $\beta$.

\subsubsection{Noisy errors}

Once logged, mass has Gaussian errors. In formulae:

\begin{eqnarray}
obslgM200_i &\sim \mathcal{N}(lgM200_i,\sigma^2_i) &\mbox{\# Gauss errors on lg mass}
\label{eqn:eq9} 
\end{eqnarray}

However, errors (as everything) is measured with a finite degree of
precision. We assume that the measured error, $obserrlgM200_i$,
is not biased (i.e. it is not systematically larger
or smaller that the true error, $\sigma_i$) but somewhat noisy.
If a $\chi^2$ distribution is adopted, it satisfies both our request of 
unbiasness and noisiness. In formulae:

\begin{eqnarray}
obserrlgM200_i^2 &\sim \sigma^2_i \chi^2_\nu / \nu &\mbox{\# Unbiased errors}
\label{eqn:eqn10}
\end{eqnarray}

where the parameter $\nu$ regulates the width of the distribution, i.e. how precise
measured errors are. Since we are 95\% confident that quoted errors are correct 
up to a factor of 2, 

\begin{eqnarray}
\nu &= 6 &\mbox{\# 95 \% confident within a factor 2}
\end{eqnarray}

\subsubsection{Prior knowledge and population structure} 

The data
used in this investigation are of quality good enough to determine all
parameters, but one, to a sufficient degree of accuracy that we should not care
about priors and we can safely
take weak (almost uniform) priors,  zeroed for un-physical values
of parameters (to avoid, for example, negative richnesses). The exception is
given by the prior on the errors (i.e. $\sigma_i$), for which there is
only one measurement per datum. The adopted prior (eq. 11) is supported by statistical
considerations (see \cite{AandH10} for details). The same prior is also used 
for the intrinsic scatter term, although any weak prior would 
return the same result, because this term is well determined by
the data.

\begin{eqnarray}
\alpha \sim& \mathcal{N}(0.0,10^4) & \mbox{\# Almost uniform prior on intercept}\\
\beta \sim& t_1 & \mbox{\# Uniform prior on angle}\\
n200_i \sim& \mathcal{U}(0,\infty) & \mbox{\# Uniform, but positive, cluster richness}\\
nbkg_i \sim& \mathcal{U}(0,\infty) & \mbox{\# Uniform, but positive, background rate} \\
1/\sigma_i^2 \sim& \Gamma(\epsilon,\epsilon) & \mbox{\# Weak prior on error}\\
1/\sigma_{scat}^2 \sim& \Gamma(\epsilon,\epsilon)& \mbox{\# Weak prior on intrinsic scatter} 
\label{eqn:eqn8}
\end{eqnarray}

Richer clusters are rarer. Therefore,
the prior on the cluster richness is, for sure, not uniform, 
contrary to our assumption (eq. 9). Modeling the population structure is
un-necessary for the data used in \cite{AandH10} and here, but is essential if
noiser richnesses were used. Indeed, \cite{AandH10} shows
that a previous published richness-mass calibration, which uses richnesses
as low as $obsn200=3$ and neglects the
$n200$ structure, shows a slope biased by five times the
quoted uncertainty. Therefore, the population structure cannot be overlooked
in general.

\subsection{Step 2: remove TeXing, perform stochastic computations and publish}

At this point, we have described, using mathematical symbols, the link between
the quantities that matter for our problem, and we only need to compute the
posterior probability distribution of the parameters by some sort of sampling (the readers
with exquisite mathematical skills may instead attempt an analytical computation). 

Just Another Gibb Sampler 
(JAGS\footnote{http://calvin.iarc.fr/$\sim$martyn/software/jags/}, 
(\cite{JAGS}) can return it at the minor cost of de-TeXing equations
1 to 12 (compare them to the JAGS code below). Poisson, Normal and Uniform distributions become
{\texttt{dpois, dnorm, dunif}}, respectively. 
JAGS, following BUGS (\cite{lunn2009bugs}), uses 
precisions, $prec = 1/\sigma^2$, in place of variances $\sigma^2$. 
Furthermore, it uses neperian logarithms, instead of decimal ones.
Eq. 5 has been rewritten using the property that the $\chi^2$
is a particular form of the Gamma distribution. 
Eq. 3 is split in two JAGS lines for a better reading.
The arrow symbol reads ``take the value of". 
{\texttt{obsvarlgM200}} is the square of $obserrlgM200$.
For computational advantages,
$\log(n200)$ is centred at an average value of 1.5 and
$\alpha$ is centred at -14.5. Finally, we replaced infinity with a large number.

The model above, when inserted in JAGS, reads:

{\footnotesize
\begin{verbatim}
model 
{
for (i in 1:length(obstot)) {
  obsbkg[i] ~ dpois(nbkg[i])                          # eq 1
  obstot[i] ~ dpois(nbkg[i]/C[i]+n200[i])             # eq 2
  n200[i] ~ dunif(0,3000)                             # eq 9
  nbkg[i] ~ dunif(0,3000)                             # eq 10
  
  precy[i] ~ dgamma(1.0E-5,1.0E-5)                    # eq 12
  obslgM200[i] ~ dnorm(lgM200[i],precy[i])            # eq 4
  obsvarlgM200[i] ~ dgamma(0.5*nu,0.5*nu*precy[i])    # eq 5

  z[i] <- alpha+14.5+beta*(log(n200[i])/2.30258-1.5)  # eq 3
  lgM200[i] ~ dnorm(z[i], prec.intrscat)              # eq 3
  }
intrscat <- 1/sqrt(prec.intrscat)                     
prec.intrscat ~ dgamma(1.0E-5,1.0E-5)                 # eq 11
alpha ~ dnorm(0.0,1.0E-4)                             # eq 7
beta ~ dt(0,1,1)                                      # eq 8
nu <-6                                                # eq 6
}
\end{verbatim}
}

JAGS samples the posterior distribution  of all quantity of interests, 
such as intercept, slope and intrinsic scatter by Gibb sampling (a sort
of Monte Carlo).
Form these samplings, it is
straightforward to compute (posterior) mean and standard deviations (by computing the
average and the standard deviation!), 
to plot posterior marginals (by ignoring the values of the other parameters)
and confidence contours, data and mean model, etc.
Therefore,  our effort is over, we only need to produce nice plots and summaries
of our results.

\begin{figure}
\centerline{\includegraphics[clip, width=8truecm]{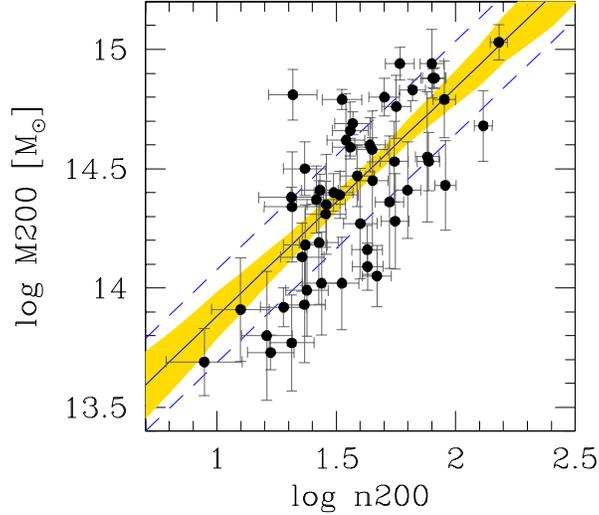}}
\caption[h]{Richness-mass scaling.
The solid line marks the mean fitted regression line, while the
dashed line shows this mean plus or minus the intrinsic scatter $\sigma_{scat}$. The shaded
region marks the 68\% highest posterior credible interval for the regression. Error bars on the
data points represent observed errors for both variables. The distances between the data and
the regression line is due in part to the measurement error and in part to the intrinsic
scatter. From \cite{AandH10}, reproduced with permission.}
\label{fig:fig1}
\end{figure}

\begin{figure*}
\centerline{\includegraphics[clip, width=16truecm]{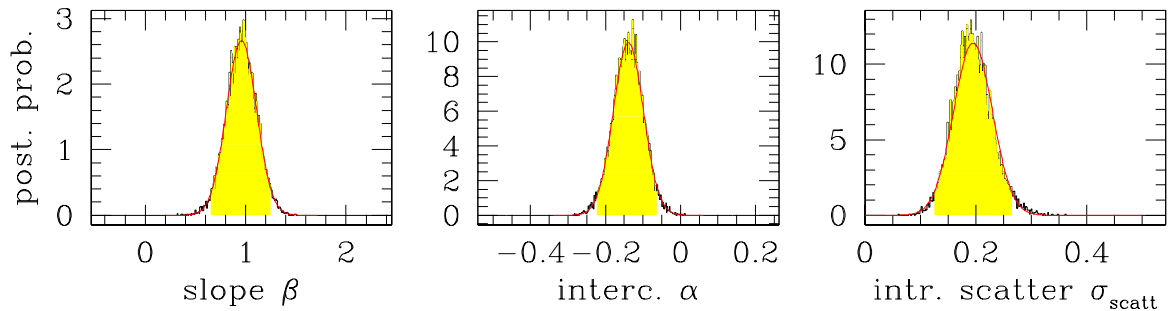}}
\caption[h]{Posterior probability distribution for the
parameters of the richness-mass scaling.
The black jagged histogram shows the posterior as computed
by MCMC, marginalised over the other parameters. The red curve
is a Gauss approximation of it.  The shaded (yellow) range shows
the 95\% highest posterior credible interval. From \cite{AandH10}, reproduced with permission.
}
\label{fig:fig2}
\end{figure*}

Figure 1 
shows the data used in this analysis (see \cite{AandH10} for details),
the mean scaling (solid line) and its 68\% uncertainty (shaded yellow
region) and the mean intrinsic scatter (dashed lines) around
the mean relation.  The $\pm 1$ intrinsic scatter band is 
not expected to contain 68\% of the data points, because of 
the presence of measurement errors. 

Figure 2 
shows the posterior marginals for
the intercept, slope and intrinsic
scatter $\sigma_{scat}$. These marginals
are reasonably well approximated by Gaussians.
The intrinsic mass scatter at a given richness, 
$\sigma_{scat}=\sigma_{lgM200|\log n200}$, is small, $0.19\pm0.03$.
(Unless otherwise stated, results of the statistical computations 
are quoted in the form $x\pm y$ where $x$
is the posterior mean and $y$ is the posterior standard deviation.)

The found relation is: 

\begin{equation}
lgM200 = (0.96\pm0.15) \ (\log n200 -1.5) +14.36\pm0.04 
\end{equation}

\subsection{Predicting masses}

As mentioned, one of the reasons why astronomers regress a quantity $x$
against another one, $y$, is to predict the latter when a direct measurement 
is missing (usually because observationally expensive to acquire).
It is clear that the uncertainty on the predicted $y$, called $\tilde{y}$ hereafter,
should account for: a) the intrinsic scatter between $y$ and $x$ (a larger
scatter implies a lower quality $\tilde{y}$ estimate); b) the uncertainty
of the $x$ (the larger it is, the noisier will be the $\tilde{y}$; c) the 
quality of the calibration between $y$ and $x$ (better determined relations should 
return more precise estimates of $\tilde{y}$); and d) extrapolation errors,
i.e. should penalize attempts to infer $\tilde{y}$ values corresponding to $x$ values absent from
the calibrator sample (e.g. outside the range sampled by it). All these requirements
are satisfied using the posterior predictive distribution,

\begin{equation} 
p( \widetilde{y} ) = \int  p( \widetilde{y} | \theta ) p(\theta| y) d \theta
\end{equation} 

where $\theta$ are the regression parameters (intercept, slope, intrinsic scatter).
This apparently ugly expression is easy to understand: one should combine (multiply)
the uncertainties of the calibrating relation, $p(\theta| y)$, to the uncertainty
of predicting new data if the calibrating relation were perfectly known, 
$p( \widetilde{y}|\theta )$. Since we are now interested in predicted values only, 
we get rid of non-interesting parameters ($\theta$) by
marginalization (integration). 

Posterior predictive distributions are so basic to be introduced at page 8 of the 
$>700$ pages ``Bayesian Data Analysis"
book (\cite{gelman2004bayesian}) and to be offered as a standard output of JAGS. Of course, we need to list
the $x$ values (richnesses), and errors of the clusters for which we want to infer $\tilde{y}$ (predicted
mass) in the data file, listing for example in the data file $obsbkg=12$, $obtot=32$, $C=5$, and
mass $obslgM200=NA$ (``not available''), to indicate that this quantity should be estimated
using the regression computed from the points with available masses and galaxy counts.
JAGS returns $p( \widetilde{y} )$ in form of sampling and therefore,
as for any other parameter, a point estimate may be obtained by taking the average and a
68 \% (credible) interval can be derived by taking the interval including 68 \% of the samplings.
Returned values behave as expected, and indeed have large errors when masses are estimated
for clusters with richnesses outside the range where the calibration has been derived
(\cite{AandM11})

\section{Second example: Cosmological parameters from SNIa}
\label{sec:SNIa}

Supernovae (SNIa) are very bright objects with very similar luminosities. The 
luminosity spread can be made even
smaller by accounting for the correlation with  
colour and stretch parameter
(the latter is a measurement of how slowly SNIa fade), as illustrated
in Figure 3 for the sample in \cite{Kessler09}.
These features make SNIa very useful
for cosmology: they can be observed far away and
the relation between flux (basically the rate of photons received) 
and luminosity (the rate of photons emitted) is modulated by the luminosity
distance (to the square), which in turns is function of the cosmological
parameters. Therefore, measuring SNIa fluxes (and redshift) allows us to put
constraints on cosmological parameters. The only minor complication is
that SNIa luminosities are function of their colour and stretch 
parameter, and these parameters have an intrinsic scatter too, 
which in turns has to be determined from the data at the same
time as the other parameters. 

\begin{figure*}
\centerline{
\subfloat[]{\includegraphics[clip, width=15truecm]{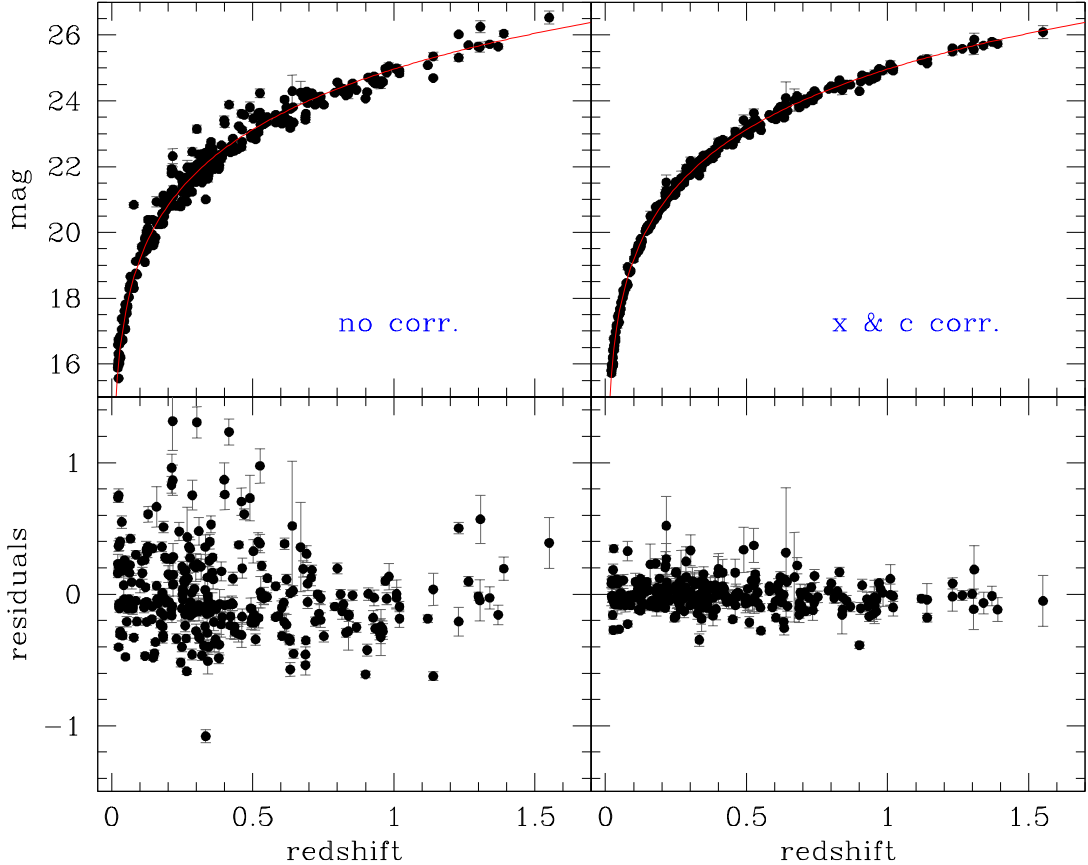}}}
\caption[h]{Apparent magnitude vs redshift of the SNIa sample (upper panels),
and their residual from a $\Omega_M=0.3$, $\Omega_\Lambda=0.7$ cosmological
model (bottom panels) before (left panels) or after (right panels) correcting
for stretching and color parameter.}
\label{fig:fig111}
\end{figure*}

\cite{March11} shows that the Bayesian approach delivers tighter statistical
constraints on the cosmological parameters over 90 \% of the times, that it reduces the
statistical bias typically by a factor $\sim 2-3$ and that it has better coverage
properties than the usual chi-squared approach. 

In this second example we can proceed a bit faster in 
illustrating this non-linear regression with
heteroscedastic errors,
non-uniform data structure and intrinsic scatter. In this example, we
also briefly discuss the prior sensitivity, i.e. how much the results
are affected by the chosen prior, and we also check the quality of the model fit.

\subsection{Step 1: put in formulae what you know}

We observe SNIa magnitudes $obsm_i$ ($=-2.5 log(flux)+c$) with Gaussian
errors $\sigma_{m,i}$, i.e.

\begin{eqnarray}
obsm_i \sim& \mathcal{N}(m_i, \sigma^2_{m,i}) 
\end{eqnarray}

\begin{figure*}
\centerline{\includegraphics[clip, width=12truecm]{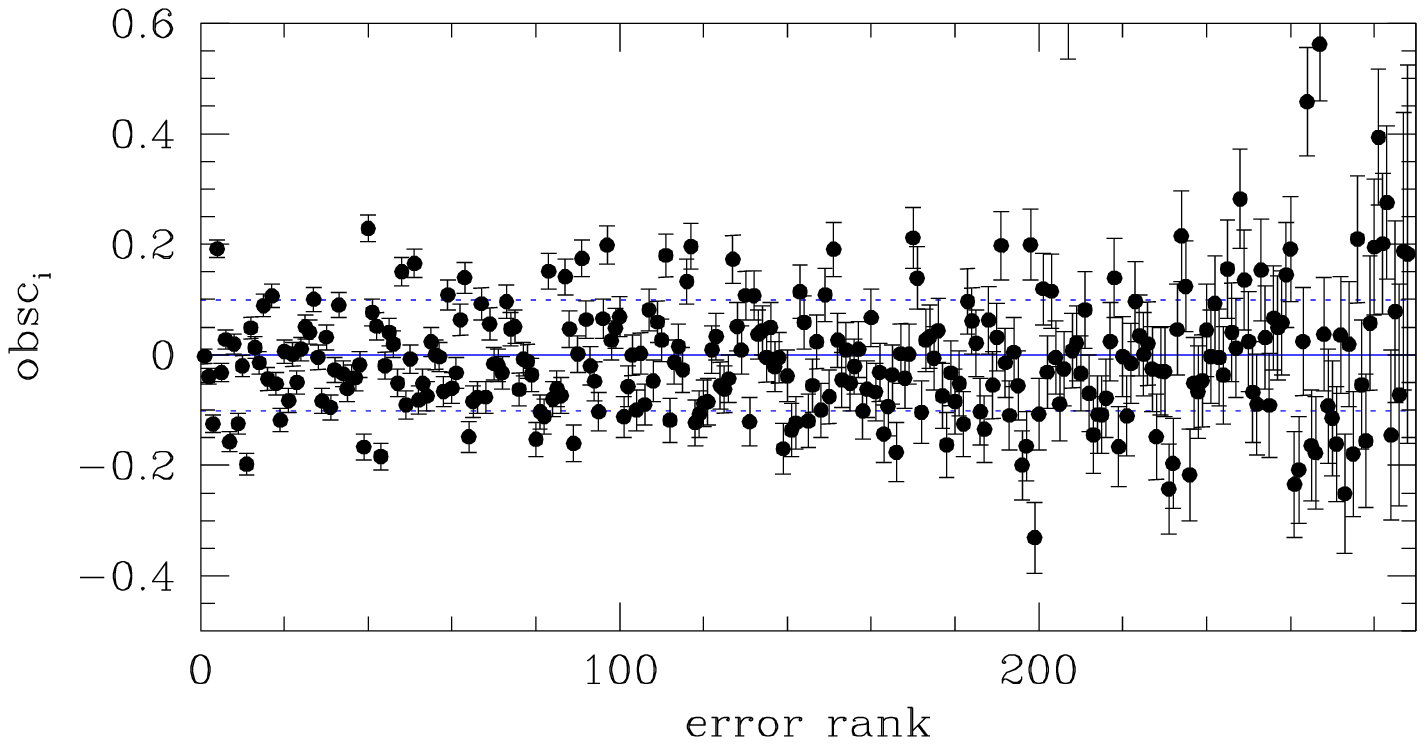}}
\centerline{\includegraphics[clip, width=12truecm]{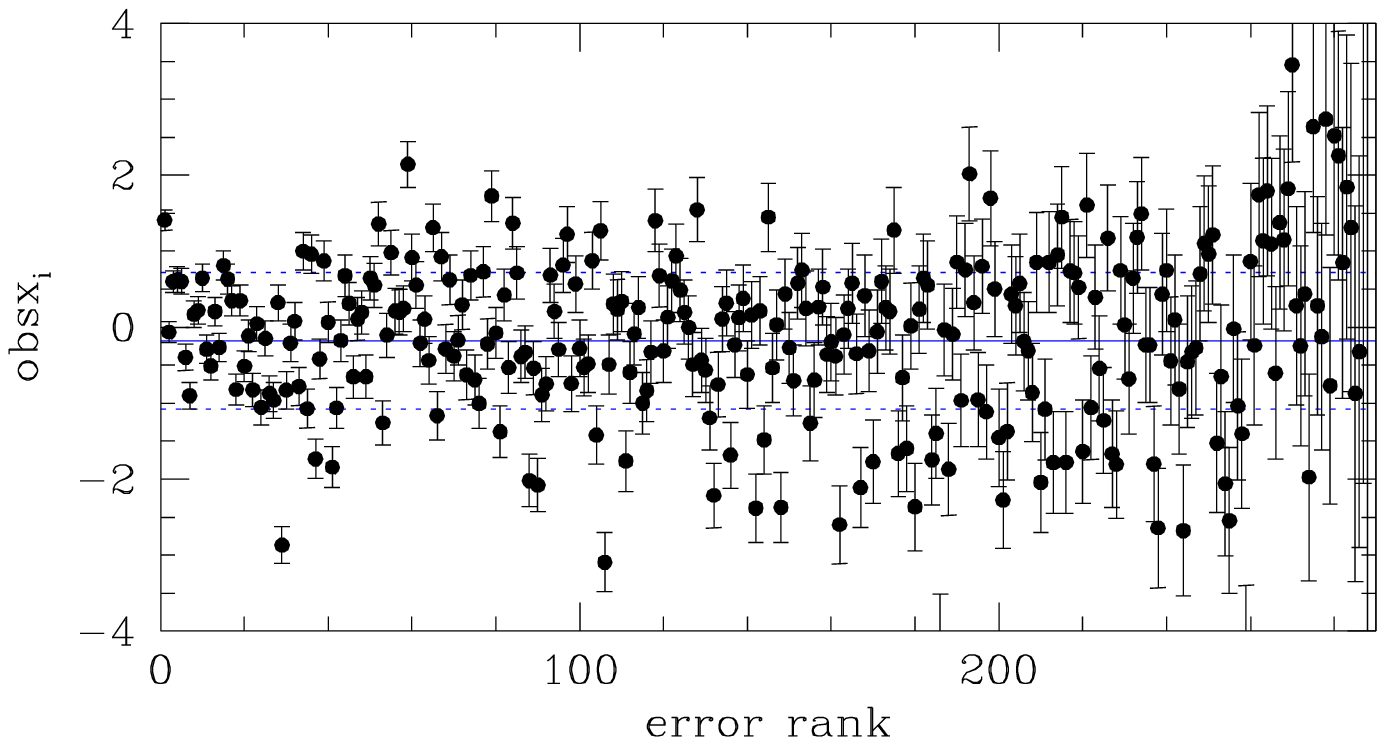}}
\caption[h]{Observed values of the stretch parameters, $obsx_i$, and
of the colour parameter, $obsc_i$ ranked by error. Points scatter
more than the error bars (see the left side of the figure). 
The dashed lines indicate the size of the intrinsic scatter as determined by
our analysis.
}
\end{figure*}

These $m_i$ are related to the distance modulus $distmod_i$,  via

\begin{eqnarray}
m_i = M+distmod_i- \alpha \, x_i + \beta \, c_i \nonumber
\end{eqnarray}

with a Gaussian intrinsic scatter $\sigma_{scat}$. More
precisely:

\begin{eqnarray}
m_i \sim& \mathcal{N}(M+distmod_i- \alpha \, x_i + \beta \, c_i, \sigma^2_{scat}) 
\end{eqnarray}

where $M$ is the (unknown) mean absolute magnitude of SNIa, 
and $\alpha$ and $\beta$ allow to reduce the SNIa luminosity scatter by 
accounting for the correlation with the stretch
and colour parameters. 

Similarly to \cite{March11},
the  $M$, $\alpha$, $\beta$  and log $\sigma_{scat}$ priors are taken uniform
in a wide range:

\begin{eqnarray}
\log_{10} \sigma_{scat} \sim& \mathcal{U}(-3, 0) \\
\alpha \sim& \mathcal{U}(-2, 2) \\
\beta \sim& \mathcal{U}(-4, 4) \\
M \sim& \mathcal{U}(-20.3, -18.3) 
\end{eqnarray}

$x_i$ and $c_i$ are the true value of the stretch and colour parameters, of which
we observe (the noisy) $obsx_i$ and $obsc_i$ with errors $\sigma_{x,i}$ and 
$\sigma_{c,i}$. In formulae:

\begin{eqnarray}
obsx_i \sim& \mathcal{N}(x_i, \sigma^2_{x,i}) \\
obsc_i \sim& \mathcal{N}(c_i, \sigma^2_{c,i}) 
\end{eqnarray}

The key point of the modeling is that the $obsx_i$ and $obsc_i$ values
scatter more than their errors, but not immensely so, see Fig 4. The presence of
a non-uniform distribution induces a Malmquist-like bias if not
accounted for (e.g. large $obsx_i$ values are more likely low $x_i$ values
scattered to large values than vice versa,
because of the larger abundance of low $x_i$ values). Therefore,
we model, as \cite{March11} do, the individual $x_i$ and $c_i$ as 
drawn from independent 
normal distributions centered on $xm$ and $cm$ with standard deviation
$R_x$ and $R_c$ respectively. In formulae:

\begin{eqnarray}
x_i \sim& \mathcal{N}(xm, R^2_{x}) \\
c_i \sim& \mathcal{N}(cm, R^2_{c}) 
\end{eqnarray}

We take uniform priors for $xm$ and $sc$, and uniform
priors on $\log R_{x}$ and on  $\log R_{c}$, between the indicated boundaries:

\begin{eqnarray}
xm \sim& \mathcal{U}(-10, +10) \\
cm \sim& \mathcal{U}(-3, +3) \\
\log_{10} R_x \sim& \mathcal{U}(-5, +2) \\
\log_{10} R_c \sim& \mathcal{U}(-5, +2) 
\end{eqnarray}

That's almost all: we need to remember the definition of distance modulus:

\begin{eqnarray}
distmod_i = 25 + 5 \log_{10} dl_i 
\end{eqnarray}

where the luminosity distance, $dl$ is a complicate expression,
involving integrals, of the redshift
$z_i$ and the cosmological parameters $\Omega_\Lambda, \Omega_M, w, H_0$
(see any recent cosmology textbook for the mathematical expression).

Redshift, in the considered sample, have heteroscedastic Gaussian
errors $\sigma_{z,i}$:

\begin{eqnarray}
obsz_i \sim& \mathcal{N}(z_i, \sigma^2_{z,i}) 
\end{eqnarray}

The redshift prior is taken uniform 

\begin{eqnarray}
z_i \sim& \mathcal{U}(0,2) 
\end{eqnarray}

Supernovae alone do not allow to determine all cosmological parameters, so
we need external prior on them, notably on $H_0$, taken from \cite{Freedman01} 
to be

\begin{eqnarray}
H_0 \sim& \mathcal{N}(72, 8^2) 
\end{eqnarray}

At this point, we may decide which sets of cosmological models we want
to investigate using SNIa, for example a flat universe with a possible
$w \ne 0$ with the following priors:

\begin{eqnarray}
\Omega_M \sim& \mathcal{U}(0, 1) \\ 
\Omega_k =&  0 \\
w \sim& \mathcal{U}(-4, 0) 
\end{eqnarray}

or a curved Universes with $w=-1$:

\begin{eqnarray}
\Omega_M \sim& \mathcal{U}(0, 1) \\
\Omega_k \sim& \mathcal{U}(-1, 0) \\
w =& -1 
\end{eqnarray}

or any other set.

Both considered cosmologies have:

\begin{eqnarray}
\Omega_k =& 1- \Omega_m - \Omega_\Lambda
\end{eqnarray}

Finally, one may want to use some data. As shortly mentioned, we use 
the compilation of 288 SNIa in \cite{Kessler09}.

\subsection{Step 2: remove TeXing, perform stochastic computations and publish}

Most of the distributions above are Normal, and the posterior
distribution can be almost completely analytically computed (\cite{March11}).
However, numerical evaluation of the stochastic part of the model
on an (obsolete) laptop takes about one minute, therefore there is no
need for speed up. Instead, the evaluation of the luminosity distance
is CPU intensive (it takes $\approx 10^3$ more
times, unless approximate analytic formulae
for the luminosity distance are used), because an integral has to be evaluated 
a number of times equal to the number of supernovae times
the number of target posterior samplings, i.e. about four millions times
in our numerical computation. 
The JAGS implementation of the luminosity distance integral
is implemented as a sum over a tightly packed grid on redshift.

As the previous example, eq 15 to 32 can be de-TeXed and used in JAGS, adding one
of the two set of priors, 33-35 or 36-38, depending on which problem one is
interested in. 

{\footnotesize
\begin{verbatim}
data {
# JAGS like precisions
 precmag <-1/errmag/errmag
 precobsc <- 1/errobsc/errobsc
 precobsx <- 1/errobsx/errobsx
 precz <- 1/errz/errz
# grid for distance modulus integral evaluation
 for (k in 1:1500){
  grid.z[k] <- (k-0.5)/1000.
 }
 step.grid.z <-grid.z[2]-grid.z[1]
}
model {
for (i in 1:length(obsz)) {
obsm[i] ~ dnorm(m[i],precmag[i])                             # eq 15
m[i] ~ dnorm(Mm+distmod[i]- alpha* x[i] + beta*c[i], precM)  # eq 16
obsc[i] ~ dnorm(c[i], precobsc[i])                           # eq 22
c[i] ~ dnorm(cm,precC)                                       # eq 24
obsx[i] ~ dnorm(x[i], precobsx[i])                           # eq 21
x[i] ~ dnorm(xm, precx)                                      # eq 23
# distmod definition & H0 term
distmod[i] <- 25 + 5/2.3026 * log(dl[i]) -5/2.3026* log(H0/300000) 
z[i] ~ dunif(0,2)                                            # eq 31
obsz[i] ~ dnorm(z[i],precz[i])                               # eq 30
######### dl computation (slow and tedious)                          
tmp2[i] <- sum(step(z[i]-grid.z) * (1+w) / (1+grid.z)) * step.grid.z
omegade[i] <- omegal * exp(3 * tmp2[i])
xx[i] <- sum(pow((1+grid.z)^3*omegam + omegade[i] + (1+grid.z)^2*omegak,-0.5)*
             *step.grid.z * step(z[i]-grid.z))
# implementing if, to avoid diving by 0 added 1e-7 to omegak
zz[1,i] <- sin(xx[i]*sqrt(abs(omegak))) * (1+z[i])/sqrt(abs(omegak+1e-7))
zz[2,i] <- xx[i]  * (1+z[i])
zz[3,i] <- (exp(xx[i]*sqrt(abs(omegak)))-exp(-xx[i]*sqrt(abs(omegak))))/2 * 
            *(1+z[i])/sqrt(abs(omegak+1e-7))
dl[i] <- zz[b,i]  
}
b <- 1 + (omegak==0) + 2*(omegak > 0)
########## end dl computation

# JAGS uses precisions
precM <- 1/ intrscatM /intrscatM
precC <- 1/ intrscatC /intrscatC
precx <- 1/ intrscatx /intrscatx
# priors
Mm~ dunif(-20.3, -18.3)                                      # eq 20
alpha ~ dunif(-2,2.0)                                        # eq 18
beta ~ dunif(-4,4.0)                                         # eq 19
cm ~ dunif(-3,3)                                             # eq 26
xm ~ dunif(-10,10)                                           # eq 25
# uniform prior on logged quantities
intrscatM <- pow(10,lgintrscatM)                             # eq 17
lgintrscatM ~ dunif(-3,0)                                    # eq 17
intrscatx <- pow(10,lgintrscatx)                             # eq 27
lgintrscatx ~ dunif(-5,2)                                    # eq 27
intrscatC <- pow(10,lgintrscatC)                             # eq 28
lgintrscatC ~ dunif(-5,2)                                    # eq 28
#cosmo priors
H0 ~ dnorm(72,1/8./8.)                                       # eq 32
omegal<-1-omegam-omegak                                      # eq 39
# cosmo priors 1st set LCDM
#omegam~dunif(0,1)                                           # eq 36
#omegak~dunif(-1,1)                                          # eq 37
#w <- -1                                                     # eq 38
# cosmo priors 2nd set: wCDM
omegam~dunif(0,1)                                            # eq 33
omegak <-0                                                   # eq 34
w ~ dunif(-4,0)                                              # eq 35
}
\end{verbatim}
}

\begin{figure*}
\centerline{
\subfloat[]{\includegraphics[clip, width=16truecm]{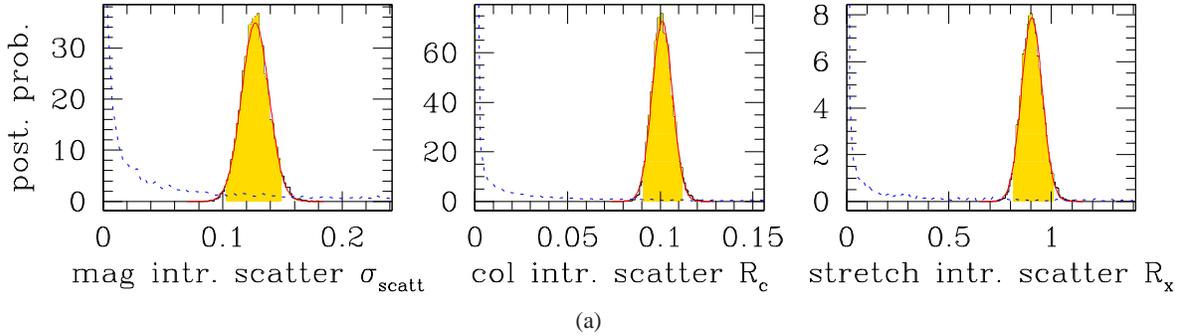}}}
\caption[h]{Prior and posterior probability distribution for the three intrinsic
scatter terms in the SNIa problem.
The black jagged histogram shows the posterior as computed
by MCMC, marginalised over the other parameters. The red curve
is a Gauss approximation of it.  The shaded (yellow) range shows
the 95\% highest posterior credible interval. The adopted priors are indicated
by the blue dotted curve. 
}
\label{fig:fig2}
\end{figure*}

Figure 5 shows the prior (dashed blue line) and posterior (histogram) probability
distribution for the three intrinsic scatter terms present in the cosmological
parameter estimation: the scatter in absolute luminosity after colour and 
stretch corrections,
($\sigma_{scat}$), the intrinsic scatter in the distribution of the colour and 
stretch terms ($R_c$ and $R_x$). This plot shows that the posterior probability
at intrinsic scatters near zero is approximately zero
and thus that the three intrinsic
scatter terms are necessary parameters for the modeling of SNIa, and not 
useless complications. 
The three posteriors are dominated by the data, being
the prior quite flat in the range where the posterior is appreciably not zero (Figure 5).
Therefore, any other prior choice, as long as smooth and shallow over the 
shown parameter range, would have returned indistinguishable results.

Not only SNIa have luminosities that
depend on colour and stretch terms, but these in turns have their own
probability distribution (taken Gaussian for simplicity) with a well
determined width. 
Figure 6 depicts the Malmquist- like bias one should incur if the spread of
the distribution of colour and 
stretch parameters is ignored: it reports the observed values (as in Fig 4),
$obsx_i$ and $obsc_i$ as well as the true values $x_i$ and $c_i$
(posterior means). The effect of equations 23 and 24
is to pull values toward the mean, and more so those with large errors,
to compensate the systematic shift (Malmquist-like bias) toward larger observed values. 


\begin{figure*}
\centerline{\includegraphics[clip, width=12truecm]{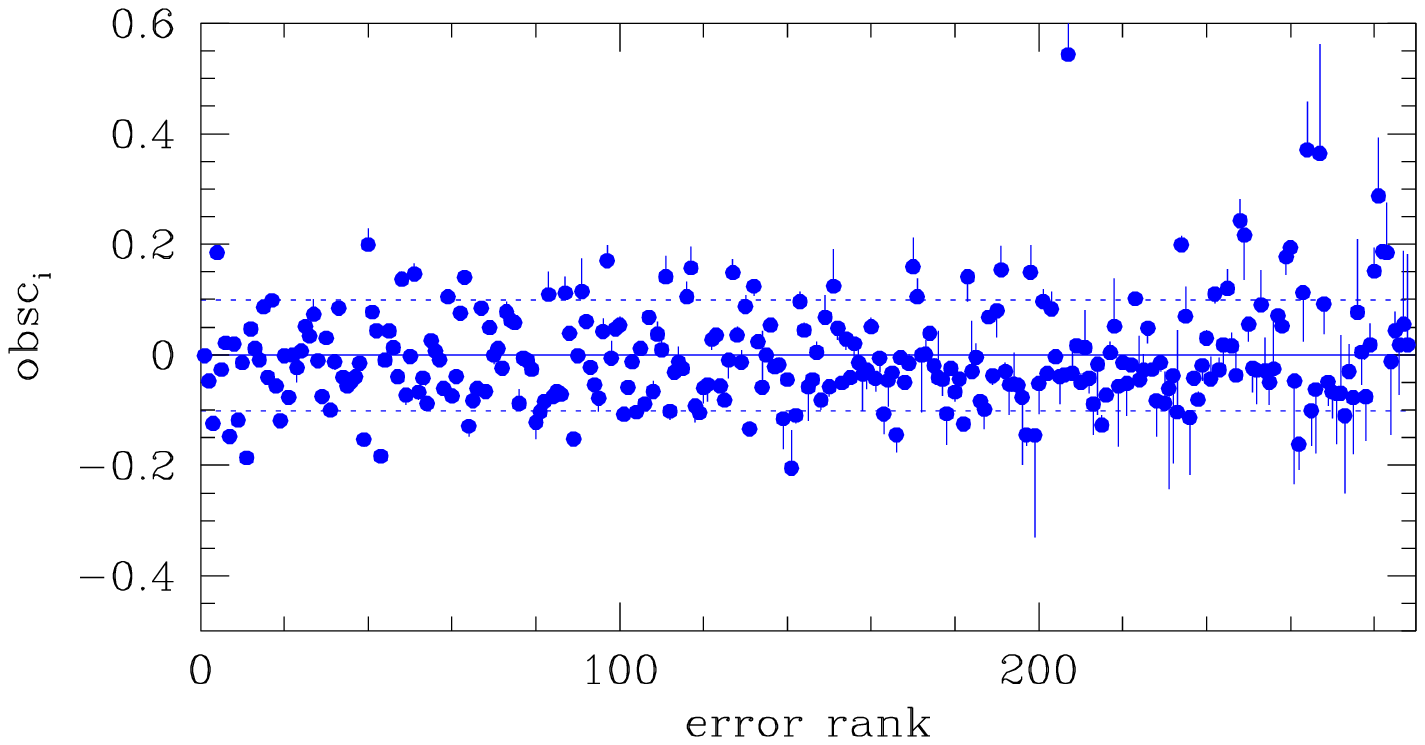}}
\centerline{\includegraphics[clip, width=12truecm]{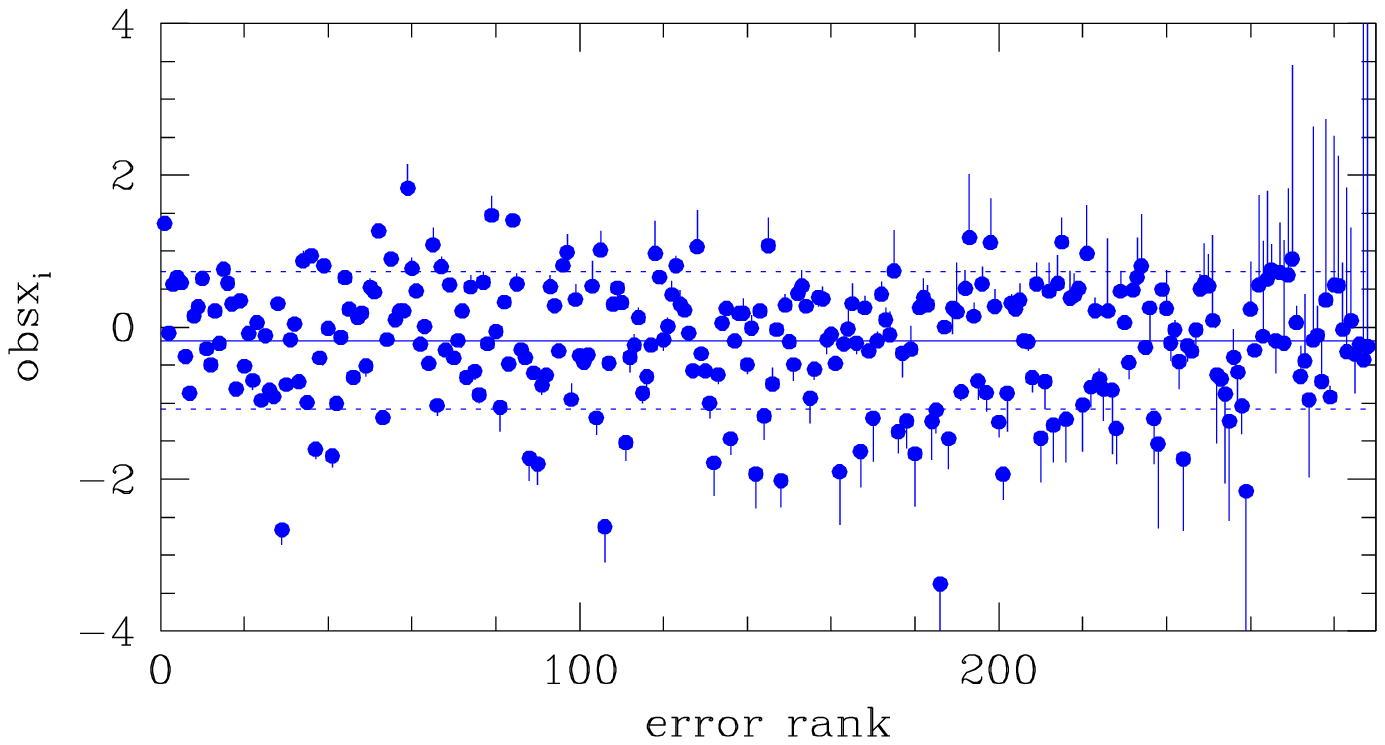}}
\caption[h]{Effect of the population structure for 
the stretch and the colour parameters. Each tick goes from the observed
value to the posterior mean.  The population modeling
attempt to counterbalance the increased spread (Malmquist-like),
especially those with larger error (on the right, in the figures), pulling
values toward the mean. 
}
\end{figure*}

Figure 7 shows the probability distribution of the two 
the colour and stretch slopes: $\alpha=0.12\pm0.02$ and $\beta=2.70\pm0.14$
respectively. As for the intrinsic scatter terms, the posterior is 
dominated by the data and therefore any other prior, smooth and shallow, 
would have returned  indistinguishable results.

Finally, Fig 8 reports perhaps the most wanted result: contours of equal probability
for the cosmological parameters $\Omega_M$ and $w$. 

For one dimensional marginals, we found: 
$\Omega_M=0.40\pm0.10$ and $w=-1.2\pm0.2$, but with non-Gaussian
probability distributions.

\begin{figure*}
\centerline{
\subfloat[]{\includegraphics[clip, width=11truecm]{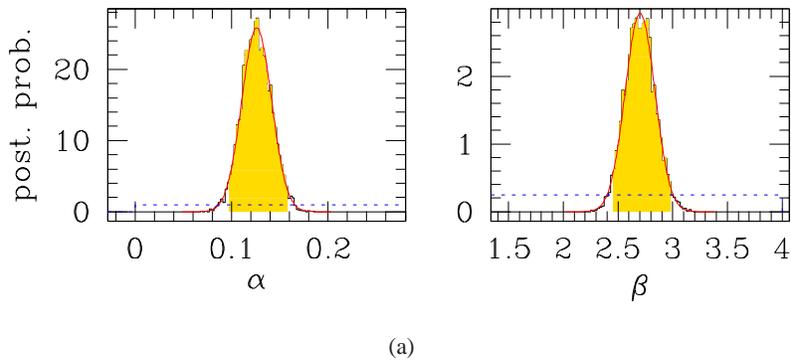}}}
\caption[h]{Prior and posterior probability distribution for the colour
and stretch slopes of the SNIa problem.
The black jagged histogram shows the posterior as computed
by MCMC, marginalised over the other parameters. The red curve
is a Gauss approximation of it.  The shaded (yellow) range shows
the 95\% highest posterior credible interval. The adopted (uniform)
priors are indicated by the blue dotted curve. 
}
\label{fig:fig2}
\end{figure*}

\begin{figure*}
\centerline{\includegraphics[clip, width=6truecm]{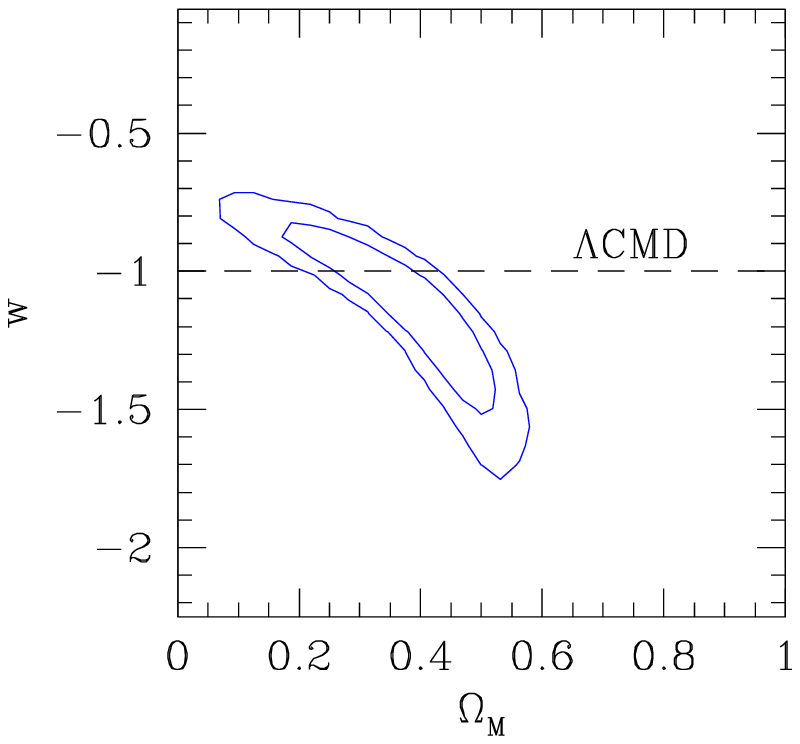}}
\caption[h]{Constraints on the cosmological parameters $\Omega_M$ and $w$.
The two contours delimit 68 \% and 95 \% constraints. 
}
\end{figure*}

\subsection{Model checking}

The work of the careful researcher does not end by finding the parameter
set that best describe the data, (s)he also
checks whether the adopted model is a good description of the data, or 
it is misspecified, i.e. in tension with the fitted data. In the non-Bayesian
paradigm this is often achieved by computing a p-value, i.e. the probability
to obtain data more discrepant than those in hand once parameters are
taken at the best fit value. The Bayesian version 
of the concept (e.g. \cite{gelman2004bayesian})
acknowledges that parameters are not perfectly known, 
and therefore one should also explore,
in addition to best fit value, other values of the parameters.  
Therefore, discrepancy becomes a vector, instead of a scalar, of dimension $j$, 
that measure the distance between the data and $j$ models, one per set of parameters
considered. Of course, 
more probable models should occur more frequently in the list to quantify
that discrepancy from an unlikely model is less detrimental than discrepancy
from a likely model.
In practice, if parameters are explored by sampling, it is just matter
of computing the discrepancy of the data in hand for each set $j$ of parameters stored
in the chain, instead of relying on one single
set of parameter (say, those that maximize the likelihood). Then, one
repeats the computation for fake data generated from the model, and counts
how many times fake data are more extreme of real data.

For example, if we want to test the modeling of the observed spread of magnitude
(i.e. equation 15 and 16), let's define:

\begin{eqnarray}
mcor_i =& M+distmod_i- \alpha \, x_i + \beta \, c_i 
\end{eqnarray}

We generate fake supernovae mag:

\begin{eqnarray}
m.fake_i \sim& \mathcal{N}(mcor_i, \sigma^2_{scat}) 
\end{eqnarray}

and fake observed values of them,

\begin{eqnarray}
obsm.fake_i \sim& \mathcal{N}(m.fake_i, \sigma^2_{m,i}) 
\end{eqnarray}

Then, we adopt a modified $\chi^2$ to quantify discrepancy (or its contrary,
agreement). For the real data set we have:

\begin{equation}
\chi^2_{real, j}= \sum_i \frac{(obsm_i-mcor_{i,j})^2}{\sigma^2_{m,i}+ E^2(\sigma_{scat})}
\end{equation}

where summation is over the data and $j$ refers to the index in the sampling chain.

Apart for the $j$ index, eq. 43 is just the usual
$\chi^2$, the difference between observed, $obsm_i$, and true $mcor_i$,  
values, weighted by the expected variance, computed as quadrature sum of errors, $\sigma_{m,i}$, and
supernovae mag intrinsic scatter $\sigma_{scat}$.
The $\chi^2$ of the $j^{th}$ fake data set, $\chi^2_{fake,j}$  is:

\begin{equation}
\chi^2_{fake,j}= \sum_i \frac{(obsm.fake_{i,j}-mcor_{i,j})^2}{\sigma^2_{m,i}+ E^2(\sigma_{scat})}
\end{equation}

At this point, we only need to compute for which fraction of
the simulations $\chi^2_{fake,j} > \chi^2_{real, j}$ and quote the result.
If the modeling is appropriate, then the computed fraction (p-value) is
not extreme (far from zero or one).
If not, our statistically modeling need to be revised, because the data
are in disagreement with the model.

We performed 15000 simulations\footnote{Skilled readers may note that we are
dealing, by large, with Gaussian distributions, and may attempt an analytic computation.}, each one generating 288 fake  
measurements of SNIa measurements. In practice, we added the following
three JAGS lines:

\begin{verbatim}
mcor[i]<-Mm+distmod[i]- alpha* x[i] + beta*c[i] # eq 40
m.fake[i] ~ dnorm(mcor, precM)                  # eq 41 
obsm.fake[i] ~ dnorm(m.fake[i],precmag[i])      # eq 42                          
\end{verbatim}

and we can simplify eq 16 in

\begin{verbatim}
m[i] ~ dnorm(mcor[i], precM)                    # eq 16  
\end{verbatim}

We found a p-value of 45 \%, i.e. that the discrepancy of the data in hand is
quite typical (similar to the one of the fake data).
Therefore, real data are quite common and
the tested part of the model shows no evidence of misspecification. The careful researcher
should then move to the other parts of the model, whose detailed
exploration is left as exercise.

\begin{figure*}
\centerline{\includegraphics[clip, width=8truecm]{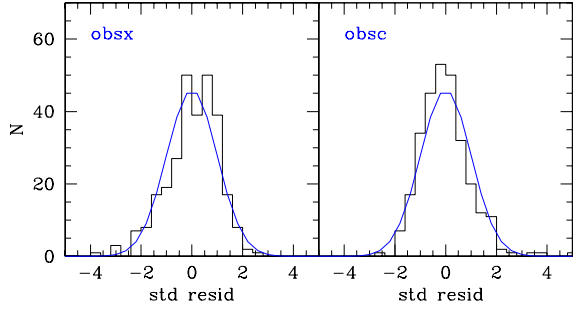}}
\caption[h]{Standardized residuals (histogram) and their expected distribution (blue curve).  
}
\end{figure*}

In such exploration of possible model misfits, it is very useful to visually inspect several
data summaries to guide the choice of which discrepancy measure one should adopt (eq. 43 or
something else?), and,
if the adopted model turns out to be unsatisfactory, to guide how to revise the modeling of the the tested 
part of the model. For example,
a possible (and common) data summary is the distribution of normalized residuals, that for $obsx_i$
reads:

\begin{equation}
stdresobsx_i = \frac{obsx_i - E(x_i)}{\sqrt{\sigma^2_{x,i}+ E^2(R_x)}}  
\end{equation}

i.e. observed minus expected value of $x_i$ divided by their expected spread 
(the sum in quadrature of errors and intrinsic spread). 
A similar summary may be built for $obsc_i$ too. To first order (at least),
standardized residuals should be normal distributed with standard deviation one
(by construction). Fig 9 shows the distribution of normalized residuals
of both $obsx_i$ and $obsc_i$, with superposed a Gaussian
centered in 0 with standard deviation equal to one (in blue). Both distributions
show a possible hint of asymmetry. At this point, the careful researcher may want to use a discrepancy
measure sensitive to asymmetries, as the skewness index, in addition to the $\chi^2$
during model testing. While leaving the actual computation to the reader,
we emphasize that if an extreme Bayesian p-value if found (on $obsx_i$ for exposing simplicity), then 
one may replace its modeling (eq. 23 in the case of $obsx_i$) with a distribution allowing a non-zero 
asymmetry and this can be easily performed in a Bayesian approach, and easily implemented
in JAGS, it is just matter of replacing the adopted Gaussian with an asymmetric distribution. 
If instead the data exploration gives an hint of double-bumped distribution 
(again on $obsx_i$ for exposing simplicity), and an extreme Bayesian p-value is found when 
a measure of discrepancy sensitive to
double-bumped distributions is adopted, then one may adopt
a mixture of Gaussians, replacing  (eq. 23 in the case of $obsx_i$) with

\begin{eqnarray}
obsx_i \sim& \lambda \mathcal{N}(x_i, \sigma^2_{x,i})+ (1-\lambda)\mathcal{N}(xx_i,
\sigma^2_{xx,i})  
\end{eqnarray}

Even more simple is the (hypothetical) case of
possible distribution (again of $obsx_i$ for exposing simplicity) with fat tails:
one may adopt a Cauchy distribution. In such case, coding in JAGS is it is just matter of replacing a \texttt{dnorm} with
\texttt{dt} in the JAGS implementation.  And so on.

In summary, model checking consists in updating the model until it produces data similar
to those in hand. One should start by
carefully and attentively inspect the data and their summaries. This
inspection should suggest a discrepancy measure to be used to quantify the model misfit and, if
one is found, to guide the model updating. The procedure should be iterated until the model 
fully reproduces the data.

\section{Summary and conclusions}
\label{sec:finalsec}

These two analysis offer a template for modeling the common awkward features
of astronomical data, namely heteroscedastic errors, non-Gaussian likelihood
(inclusive of upper/lower limits), non-uniform populations or data structure,
intrinsic scatter (either due to unidentified source of errors, or due
to population spreads), noisy estimates of the errors, 
mixtures and prior knowledge.

In a Bayesian framework, learning from the data  and the prior it is just
matter of formalizing in mathematical terms our wordy
statements about the quantities under investigation and how the data arize.
The actual numerical computation of the posterior probability distribution of the
parameters is left to (special) Monte Carlo programs, of which we don't need
to be afraid more than numerical methods (Monte Carlo) used to compute
the integral of a function.

The great advantage of the Bayesian modeling is its high flexibility: 
if the data (or theory) call for a more complex modeling, or call 
for using distributions
different from those initially taken, it is just matter of replacing them
in the model, because there are no simplifications forced by the need of reaching 
the finishing line, as instead is the case in other modeling approaches. Furthermore,
if  one is interested in constraining another cosmological model,
for example one with a redshift-dependent dark energy equation of state  
$w = w_0+w_1(1+z)/(1+z_p)$, one should just replace $w$ in the JAGS code 
with the above equation. This flexible, hierarchical, modeling
is native with Bayesian methods. 

In both our two examples, the Bayesian approach performs, unsurprisingly, better than 
than non-Bayesian methods obliged to discard part of the available
information in order to reach the finishing line:
Bayesian methods just use all the provided information. 

As a final note, we remember that
the careful researcher, whether using a Bayesian modeling or not,
before publishing his own result should check that the
numerical computation is adequate for his purpose and that the model is
appropriate for the data. Therefore, 
if you, gentle reader, are using our examples as template, 
remember to include in your work a sensitivity
analysis, by checking that your assumptions (both likelihood and
prior) are reasonable. Some prior sensitivity analysis has been performed
in both examples to emphasize the importance of showing how much conclusions 
(the posterior) rely on assumptions (prior). For what concerns
the likelihood, i.e. misfitting, the key point consists in updating the model 
fomulation until it produces data similar
to those in hand, as we have shown in great detail for the second example.

More in general, we hope that the template modeling shown in these two examples
may be useful for any analysis confronted with 
modeling the awkward features of astronomical 
data, among which heteroscedastic (point-dependent) 
errors, intrinsic scatter, data structure, non-uniform population (often
called Malmquist bias) and non-Gaussian data, inclusive of upper/lower limits.

\medskip
{\it Acknowledgements:}
The first part of this chapter is largely based on papers written in collaboration
with Merrilee Hurn. It is a pleasure to thank Merrilee for the fruitful collaboration
and her wise suggestions along the years, and for comments on an early draft of this
chapter. Errors and inconsistencies remain my own.

\bibliographystyle{unsrt}
\bibliography{springer}

\end{document}